\documentclass{article}
\usepackage{graphicx, amssymb, latexsym}

\def \a{\alpha}
\def \b{\beta}
\def \g{\gamma}
\def \d{\delta}

\def \z{\zeta}

\def \l{\lambda}
\def \m{\mu}
\def \n{\nu}

\def \ph{\phi}

\def \c{\chi}
\def \ps{\psi}
\def \om{\omega}

\def \La{\Lambda}

\def \Ph{\Phi}
\def \Ps{\Psi}

\def \la#1{\label{#1}}
\def \ift{\infty}
\def \le{\left}
\def \ri{\right}
\def \da{\dagger}
\def \ti#1{\tilde{#1}}
\def \lb{\lbrack}
\def \rb{\rbrack}

\def \rar{\rightarrow}

\def \ld{\ldots}
\def \cd{\cdots}
\def \nn{\nonumber}

\newcommand \beq{\begin{eqnarray}}
\newcommand \eeq{\end{eqnarray}}
\newcommand \bea{\begin{eqnarray*}}
\newcommand \eea{\end{eqnarray*}}
\newcommand \ben{\begin{enumerate}}
\newcommand \een{\end{enumerate}}
\newcommand \ba{\begin{array}}
\newcommand \ea{\end{array}}

\def \Tr{{\rm Tr}}

\begin{document}

\begin{center}
   {\LARGE\bf Multicritical Matrix-Vector Models of} \\
   \vspace{.3cm}
   {\LARGE\bf Quantum Orbifold Geometry} \\
   \vspace{.7cm}
   {\large {\bf C.-W. H. Lee}\footnote
           {e-mail address: h11lee@scimail.uwaterloo.ca}} \\
   \vspace{.7cm}
   {\it Department of Physics, Faculty of Science, University of 
   Waterloo, Waterloo, Ontario, Canada, N2L 3G1.} \\
   \vspace{.4cm}
   {\large October 14, 2003} \\
   \vspace{.7cm}
   {\large\bf Abstract}
\end{center}

\noindent
We construct bosonic and fermionic matrix-vector models which describe 
orbifolded string worldsheets at a limit in which the dimension of the vector 
space and the matrix order are taken to infinity.  We evaluate tree-level 
one-loop or multiloop amplitudes of these string worldsheets by means of
Schwinger--Dyson equations and derive their expressions at the multicritical 
points.  Some of these amplitudes resemble or are closely related to those of 
ordinary multicritical Hermitian matrix models by a constant factor, whereas 
some differ significantly.

\vspace{.5cm}

\begin{flushleft}
\end{flushleft}
\pagebreak

\section{Introduction}
\la{s1}

Large-$N$ matrix models provide us with valuable insights into 
non-perturbative behavior of low-dimensional bosonic strings.  (See 
Ref.~\cite{dgz} and the references therein.)  This is rendered possible by the 
observation that the dual of Feynman diagrams of these models may be regarded as 
discretised oriented string worldsheets and by the tractability of these models 
at the double scaling limit.  Recent work has revealed that these models are 
well suited to the study of D-brane dynamics \cite{mv, dkkmms}, too.

There are other important string models besides oriented string theory.  For 
instance, one may construct type~I superstring theory by an orientifold
projection of type~IIB theory \cite{ps, horava, schwarz}.  The worldsheets
involved are orbifolded and respect a ${\mathbb Z}_2$ symmetry which 
interchanges left- and right-movers.  Recently, we have discovered a family of 
matrix-vector models which not only serve as examples of noncommutative
probability of type~B \cite{bgn} 
but also may be used to study models of orbifolded string 
worldsheets \cite{0303086}.  The basic ingredients of these models are vectors 
of square matrices of Grassmann numbers.  If both the vector dimension and the 
order of the matrices are, loosely speaking, taken to infinity, then the Feynman 
diagrams are the dual of discretised orbifolded string worldsheets.  It is 
possible to evaluate the tree-level one-loop amplitudes of the simplest of these 
models.  It would certainly be of interest if the calculations can be extended 
to multiloop amplitudes of multicritical matrix-vector models.  Such 
calculations are the subject matter of this article.

Moreover, we will show that there are bosonic counterparts to these fermionic
models.  We will see that the orbifolded string worldsheets that are contructed
from the bosonic models display some unique characteristics.

Here is a brief synopsis of this article.  In Section~\ref{s2}, we will 
introduce bosonic and fermionic matrix-vector models which describe string
worldsheets homeomorphic to ${\mathbb R}^2 / {\mathbb Z}_2 \times {\mathbb Z}_2$.
We will derive the tree-level multiloop amplitudes at the multicritical points 
via Schwinger--Dyson equations.  In Section~\ref{s3}, we will turn our attention
to models which describe string worldsheets homeomorphic to ${\mathbb R}^2 /
{\mathbb Z}_2$ and use a similar method to evaluate the tree-level one-loop
amplitudes at the multicritical points.  Then we will summarise our results and
point out future directions of this work in Section~\ref{s4}.

\section{Multicritical models of ${\mathbb R}^2 / {\mathbb Z}_2 \times 
{\mathbb Z}_2$} 
\la{s2}

Consider a fermionic matrix-vector model whose building blocks are  
Grassmann matrices $\Ps_{\m}$ and $\bar{\Ps}_{\m}$ of order $N_m$, where $\m$
may take any integer value between 1 and $N_v$ inclusive and is called a 
vector index. The action of the model takes the form
\beq
   \lefteqn{S_f := N_m \sqrt{N_v} \sum_{\m = 1}^{N_v} \Tr \bar{\Ps}_{\m} 
   \Ps_{\m} - \frac{N_m^2 (g_1 - 1)}{2} \sum_{\m_1, \m_2 = 1}^{N_v} \le\lb \Tr 
   \le( \bar{\Ps}_{\m_1} \Ps_{\m_2} \ri) \ri\rb^2 } \nn \\
   & & - N_m \sum_{n=1}^{\ift} \frac{c_n}{2n}
   \sum_{\m_1, \m_2, \ld, \m_{2n} = 1}^{N_v} \Tr \le\lb \le(
   \bar{\Ps}_{\m_1} \Ps_{\m_2} \bar{\Ps}_{\m_3} \Ps_{\m_4} \cd
   \bar{\Ps}_{\m_{2n-1}} \Ps_{\m_{2n}} \ri)^2 \ri\rb \nn \\
   & & - N_m^2 \sum_{n=2}^{\ift} \frac{g_n}{2n}
   \sum_{\m_1, \m_2, \ld, \m_{2n} = 1}^{N_v} \le\lb \Tr \le(
   \bar{\Ps}_{\m_1} \Ps_{\m_2} \bar{\Ps}_{\m_3} \Ps_{\m_4} \cd
   \bar{\Ps}_{\m_{2n-1}} \Ps_{\m_{2n}} \ri) \ri\rb^2,
\la{2.1}
\eeq
where $c_n$ and $g_n$ are constant complex numbers for $n$ = 1, 2, 3, \ld, 
and so on.  Like the models we studied in Ref.~\cite{0303086}, the dual of the 
Feynman diagrams of this model in the double large-$N$ limit in which we
take $N_v$ to infinity first and $N_m$ to infinity afterwards may be 
identified as quadrangulated surfaces of the orbifold ${\mathbb R}^2 / 
{\mathbb Z}_2 \times {\mathbb Z}_2$.  Note that the expression 
\[ - \frac{N_m}{2} \sum_{\m_1, \m_2 = 1}^{N_v} \le\lb \Tr \le(
   \bar{\Ps}_{\m_1} \Ps_{\m_2} \ri) \ri\rb^2 \]
in Eq.~(\ref{2.1}) may be represented as a pair of Feynman propagators.  The 
term
\[ - \frac{N_m^2 (g_1 - 1)}{2} \sum_{\m_1, \m_2 = 1}^{N_v} \le\lb \Tr 
   \le( \bar{\Ps}_{\m_1} \Ps_{\m_2} \ri) \ri\rb^2 \]
is put into $S_f$ for future convenience.  

Let
\beq
   Z_f (N_m, N_v) := \int d\Ps_1 d\bar{\Ps}_1 d\Ps_2 d\bar{\Ps}_2 \cd
   d\Ps_{N_v} d\bar{\Ps}_{N_v} \exp S_f
\la{2.1a}
\eeq
be the partition function of this model.  The quantities which are of interest
to us are the connected Green function
\beq
   \lefteqn{G_f (p_1, p_2, \ld, p_{\ti{n}}; k_1, k_2, \ld, k_n) := 
   \lim_{N_m \rar \ift} \lim_{N_v \rar \ift} N_m^{\ti{n} + 2n - 2}} \nn \\ 
   & & \sum_{\n_{1, 1}, \n_{2, 1}, \ld, \n_{2 p_1, 1} = 1}^{N_v}
   \sum_{\n_{1, 2}, \n_{2, 2}, \ld, \n_{2 p_2, 2} = 1}^{N_v} 
   \cd \sum_{\n_{1, \ti{n}}, \n_{2, \ti{n}}, \ld, \n_{2 p_{\ti{n}}, 
   \ti{n}} = 1}^{N_v} \nn \\ 
   & & \sum_{\m_{1, 1}, \m_{2, 1}, \ld, \m_{2 k_1, 1} = 1}^{N_v}
   \sum_{\m_{1, 2}, \m_{2, 2}, \ld, \m_{2 k_2, 2} = 1}^{N_v} \cd
   \sum_{\m_{1, n}, \m_{2, n}, \ld, \m_{2 k_n, n} = 1}^{N_v} \nn \\
   & & \le\langle \prod_{j=1}^{\ti{n}}
   \Tr \le\lb \le( \bar{\Ps}_{\n_{1, j}} \Ps_{\n_{2, j}} 
   \bar{\Ps}_{\n_{3, j}} \Ps_{\n_{4, j}} \cd \bar{\Ps}_{\n_{2p_j - 1, j}} 
   \Ps_{\n_{2p_j, j}} \ri)^2 \ri\rb \ri. \nn \\
   & & \le. \cdot \prod_{i=1}^n
   \le\lb \Tr \le( \bar{\Ps}_{\m_{1, i}} \Ps_{\m_{2, i}} 
   \bar{\Ps}_{\m_{3, i}} \Ps_{\m_{4, i}} \cd 
   \bar{\Ps}_{\m_{2k_i - 1, i}} \Ps_{\m_{2k_i, i}} \ri) \ri\rb^2
   \ri\rangle_{{\rm conn}, S_f},
\la{2.2} 
\eeq
where $n$ is any non-negative integer, $\ti{n}$ is any positive integer, $p_1$, 
$p_2$, \ld, $p_{\ti{n}}$, $k_1$, $k_2$, \ld, and $k_n$ are also any positive 
integers, and the subscripts "conn" and $S_f$ tell us that this Green function 
is connected and that the expectation value is evaluated with respect to the 
action $S_f$, respectively.  Terms of some examples of Green functions are 
depicted in Fig.~\ref{f1}.

\begin{figure}
\centering
\includegraphics[width=2.9in, height=6.7in]{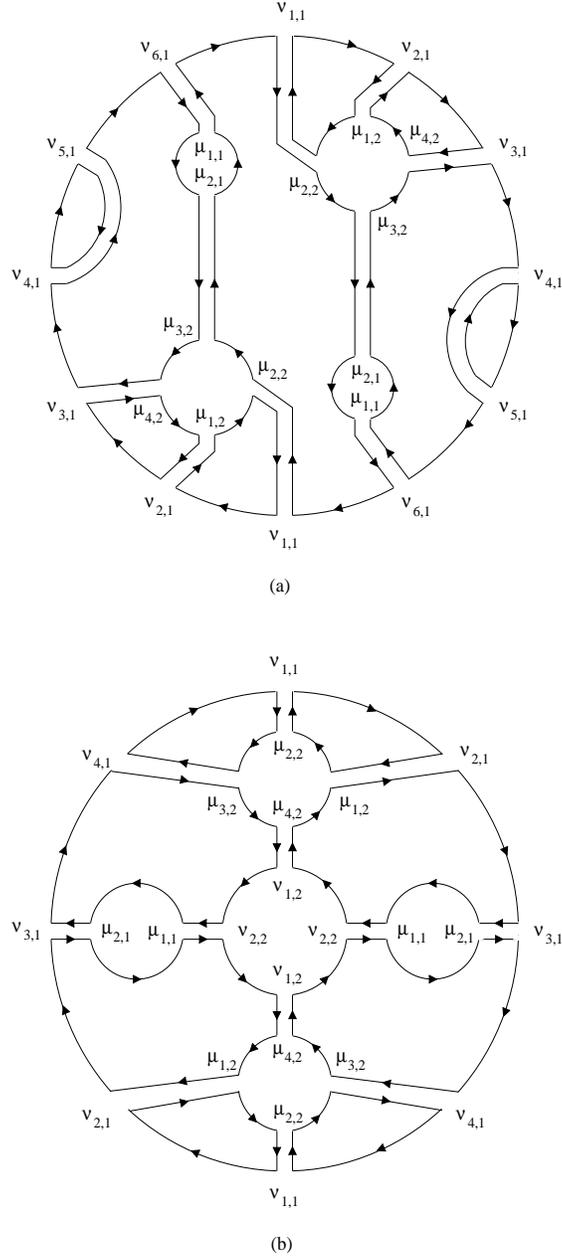}
\caption{\em (a) A Feynman diagram of $G_s (3; 1, 2)$, where $s = f$ or $b$.  
This diagram corresponds to a vanishing term if $s = f$.  (b) A Feynman 
diagram of $G_s (2, 1; 1, 2)$.  In both diagrams, the indices $\m_{i,j}$ or 
$\n_{i,j}$, where $i$ and $j$ are positive integers, are vector indices.}
\la{f1}
\end{figure}

There is a bosonic counterpart to the fermionic model.  Let $M_1$, $M_2$, \ld, 
and $M_{N_v}$ be complex matrices of order $N_m$.  Out of these matrices may 
be constructed a bosonic matrix-vector model whose action is
\bea
   \lefteqn{S_b := - N_m \sqrt{N_v} \sum_{\m = 1}^{N_v} \Tr M^{\da}_{\m} 
   M_{\m} - \frac{N_m^2 (g_1 - 1)}{2} \sum_{\m_1, \m_2 = 1}^{N_v} \le\lb \Tr 
   \le( M^{\da}_{\m_1} M_{\m_2} \ri) \ri\rb^2 } \\
   & & - N_m \sum_{n=1}^{\ift} \frac{c_n}{2n}
   \sum_{\m_1, \m_2, \ld, \m_{2n} = 1}^{N_v} \Tr \le\lb \le(
   M^{\da}_{\m_1} M_{\m_2} M^{\da}_{\m_3} M_{\m_4} \cd
   M^{\da}_{\m_{2n-1}} M_{\m_{2n}} \ri)^2 \ri\rb \\
   & & - N_m^2 \sum_{n=2}^{\ift} \frac{g_n}{2n}
   \sum_{\m_1, \m_2, \ld, \m_{2n} = 1}^{N_v} \le\lb \Tr \le(
   M^{\da}_{\m_1} M_{\m_2} M^{\da}_{\m_3} M_{\m_4} \cd
   M^{\da}_{\m_{2n-1}} M_{\m_{2n}} \ri) \ri\rb^2.
\eea
Unlike the fermionic model, those Feynman diagrams of this bosonic model in 
which there is no vertex representing a term whose coefficient is $c_i$, 
where $i$ is any positive integer, do not vanish.  Such non-zero Feynman 
diagrams are also invariant under parity transformation.  Let
\beq
   Z_b (N_m, N_v) := \int dM^{\da}_1 dM_1 dM^{\da}_2 dM_2 \cd 
   dM^{\da}_{N_v} dM_{N_v} \exp S_b 
\la{2.3}
\eeq
be the partition function of this model.  The physical quantities we would 
like to evaluate are the connected Green functions 
\[ G_b (p_1, p_2, \ld, p_{\ti{n}}; k_1, k_2, \ld, k_n) \]
defined as in Eq.~(\ref{2.2}) with $\bar{\Ps}$, $\Ps$ and the subscript 
$S_f$ replaced with $M^{\da}$, $M$, and the subscript $S_b$, respectively.

\subsection{Schwinger--Dyson equations}
\la{s2.1}

We may evaluate the multiloop amplitudes of these matrix-vector models by 
means of Schwinger--Dyson equations.  The results are intimately related to 
the ordinary Hermitian matrix model whose action is
\[ S_H := - N_m \Tr V (\Ph), \] 
where $\Ph$ is a Hermitian matrix of order $N_m$ and
\[ V (\Ph) := \sum_{n=1}^{\ift} \frac{g_n}{2n} \Ph^{2n}. \]
Let
\beq
   \ti{\ph} (n) := \lim_{N_m \rar \ift} \frac{1}{N_m} \langle 
   \Tr \, \Ph^{2n} \rangle_{S_H} 
\la{2.3a}
\eeq
be the expectation value of $\Tr \, \Ph^{2n}$.  Consider the trivial equations
\bea
   \lefteqn{\lim_{N_m \rar \ift} \lim_{N_v \rar \ift} 
   \frac{1}{N_m^2 \sqrt{N_v} Z_f (N_m, N_v)}} \\
   & & \cdot \sum_{i,j=1}^{N_m} \sum_{\a_0 = 1}^{N_v} \int d\Ps_1 
   d\bar{\Ps}_1 d\Ps_2 d\bar{\Ps}_2 \cd d\Ps_{N_v} d\bar{\Ps}_{N_v} \\
   & & \frac{\partial}{\partial \bar{\Ps}_{\a_0 ij}} \{
   \sum_{\a_1, \a_2, \ld, \a_{2n-1} = 1}^{N_v}
   \le( \bar{\Ps}_{\a_1} \Ps_{\a_2} \cd \bar{\Ps}_{\a_{2n-1}} \Ps_{\a_0} 
   \bar{\Ps}_{\a_1} \Ps_{\a_2} \cd \bar{\Ps}_{\a_{2n-1}} \ri)_{ij} \\
   & & \exp S_f \} = 0 
\eea
and
\bea
   \lefteqn{\lim_{N_m \rar \ift} \lim_{N_v \rar \ift}
   \frac{1}{N_m^2 \sqrt{N_v} Z_b (N_m, N_v)}} \\
   & & \cdot \sum_{i,j=1}^{N_m} \sum_{\a_0 = 1}^{N_v} \int dM^{\da}_1 dM_1 
   dM^{\da}_2 dM_2 \cd dM^{\da}_{N_v} dM_{N_v} \\
   & & \frac{\partial}{\partial M^{\da}_{\a_0 ij}} \{
   \sum_{\a_1, \a_2, \ld, \a_{2n-1} = 1}^{N_v}
   \le( M^{\da}_{\a_1} M_{\a_2} \cd M^{\da}_{\a_{2n-1}} M_{\a_0} 
   M^{\da}_{\a_1} M_{\a_2} \cd M^{\da}_{\a_{2n-1}} \ri)_{ij} \\
   & & \exp S_b \} = 0,
\eea
where $n$ is an arbitrary positive integer, for the fermionic and bosonic models, 
respectively.  They yield the Schwinger--Dyson equation (see Ref.~\cite{0303086} 
for some intermediate steps),
\beq
   \lefteqn{2 \sum_{k=0}^{n-2} \ti{\ph} (k) G_s (n-1-k) + 
   2 \d_{sb} \ti{\ph} (n-1)} \nn \\
   & & - \sum_{k=1}^{\ift} c_k \ti{\ph} (n+k-1) - 
   \sum_{k=1}^{\ift} g_k G_s (n+k-1) = 0, 
\la{2.5}
\eeq
where $s = f$ or $b$, and $\d_{sb}$ (or $\d_{sf}$) is a Kronecker delta function.
The first sum vanishes if $n=1$.  Define
\[ \om_s (\z) := \sum_{n=1}^{\ift} \frac{G_s (n)}{\z^{2n+1}} \]
and
\[ \c (\z) := \sum_{n=1}^{\ift} \frac{\ti{\ph} (n)}{\z^{2n+1}} \]
as the spectral density functions of the matrix-vector models and of the
Hermitian matrix model, respectively.  Then Eq.~(\ref{2.5}) leads to
\bea
   \lefteqn{\om_s (\z) = \le\{ - \d_{sf} \frac{2}{\z^2} \c (\z) 
   - \sum_{k=1}^{\ift} c_k \z^{2k-2} \c (\z) \ri.} \\
   & & \le. + \sum_{k=0}^{\ift} \z^{2k-1} 
   \sum_{l=0}^{\ift} \le\lb \ti{\Ph} (l) c_{k+l+1} + G (l) g_{k+l+1} 
   \ri\rb \ri\} / \le\lb \sum_{k=1}^{\ift} g_k \z^{2k-2} -
   \frac{2}{\z} \c (\z) \ri\rb.
\eea
It is well known (see, e.g., Ref.~\cite{ajm}) that
\[ \c (\z) = \frac{1}{2} \le\lb V' (\z) - M(\z, \b) \sqrt{\z^2 - \b} \ri\rb, \]
where $\b$ is determined by the integral relation
\beq
   W(\b) := - \frac{2}{\pi i} \int_0^{\sqrt{\b}} 
   \frac{q V' (q) dq}{\sqrt{q^2 - \b}} = 2,
\la{2.5a}
\eeq
and
\beq
   M(\z, \b) := \sum_{j=1}^{\ift} \z^{2j-2} \sum_{k=0}^{\ift}
   \frac{(2k)!}{4^k (k!)^2} g_{k+j} \b^k. 
\la{2.5b}
\eeq
Hence 
\beq
   \om_s (\z) = \frac{1}{2} \le( \frac{2}{\z} \d_{sf} +
   \sum_{k=1}^{\ift} c_k \z^{2k-1} \ri) + \frac{P_s (\z)}{\sqrt{\z^2 - \b}}, 
\la{2.6}
\eeq
where $P_s (\z)$ is a polynomial.

$P_s (\z)$ may be determined by the holomorphic properties of $\om_s (\z)$.  
Multiplying both sides of Eq.~(\ref{2.6}) by $\pm i \sqrt{(\sqrt{\b} - \z)
(\sqrt{\b} + \z)}$, we obtain
\bea
   \lefteqn{\pm \om_s (\z \pm i0) i \sqrt{(\sqrt{\b} - \z) 
   (\sqrt{\b} + \z)} = } \\
   & & \pm \frac{1}{2} \le( \frac{2}{\z} \d_{sf} +
   \sum_{k=1}^{\ift} c_k \z^{2k-1} \ri) i \sqrt{(\sqrt{\b} - \z) 
   (\sqrt{\b} + \z)} + P_s (\z). 
\eea
Thus we get a discontinuity equation
\bea
   \lefteqn{\le\lb \om_s (\z + i0) + \om_s (\z - i0) \ri\rb
   i \sqrt{(\sqrt{\b} - \z) (\sqrt{\b} + \z)} = } \\
   & & \le( \frac{2}{\z} \d_{sf} + \sum_{k=1}^{\ift} c_k \z^{2k-1} \ri) 
   i \sqrt{(\sqrt{\b} - \z) (\sqrt{\b} + \z)}. 
\eea
As a result,
\beq 
   \om_s (\z) = \frac{1}{2 \pi \sqrt{\z^2 - \b}} \int_{- \sqrt{\b}}^{\sqrt{\b}}
   \frac{ d\l \sqrt{\b - \l^2}}{\l - \z} \le( \frac{2}{\l} \d_{sf} + 
   \sum_{k=1}^{\ift} c_k \l^{2k-1} \ri) + \frac{Q_s (\z)}{\sqrt{\z^2 - \b}}, 
\la{2.7}
\eeq
where $Q_s (\z)$ is a polynomial.  Since $\lim_{\z \rar \ift} \om_s (\z) = 0$,
$Q_s (\z)$ is constant.  Then $\lim_{\z \rar \ift} \z \om_s (\z) = 1$ implies 
that 
\[ Q_s (\z) \equiv 1. \]
Evaluating the integral in Eq.~(\ref{2.7}) and comparing the result with 
Eq.~(\ref{2.6}) then imply
\beq
   P_s (\z) = 2 \d_{sb} - \sum_{n=1}^{\ift} c_n \z^{2n} +
   2 \sum_{n=0}^{\ift} \z^{2n} \sum_{k=1}^{\ift} c_{k+n} \b^k
   \frac{(2k-2)!}{4^k k! (k-1)!}. 
\la{2.8}
\eeq
We may use Eqs.~(\ref{2.6}) and (\ref{2.8}) to expand $\om_s (\z)$ as a power 
series in $1/\z$ and obtain all connected Green functions of the form 
$G_s (p)$.

\subsection{Multiloop correlators}
\la{s2.2}

To obtain other connected Green functions, we apply the formula
\beq
   \lefteqn{G_s (p_1, p_2, \ld, p_{\ti{n}}; k_1, k_2, \ld, k_n)} \nn \\
   & = & -2 p_{\ti{n}} \frac{\partial}{\partial c_{p_{\ti{n}}}} 
   G_s (p_1, p_2, \ld, p_{\ti{n} - 1}; k_1, k_2, \ld, k_n) \nn \\
   & = & -2 k_n \frac{\partial}{\partial g_{k_n}}
   G_s (p_1, p_2, \ld, p_{\ti{n}}; k_1, k_2, \ld, k_{n-1}). 
\la{2.9}
\eeq
Let
\bea
   \lefteqn{\om_s (\z_1, \z_2, \ld, \z_{\ti{n}}; z_1, z_2, \ld, z_n) :=} \\
   & & \sum_{p_1, p_2, \ld, p_{\ti{n}} = 1}^{\ift}
   \sum_{k_1, k_2, \ld, k_n = 1}^{\ift} 
   \frac{G_s (p_1, p_2, \ld, p_{\ti{n}}; k_1, k_2, \ld, k_n)}    
   {\z_1^{2 p_1 + 1} \z_2^{2 p_2 + 1} \cd \z_{\ti{n}}^{2 p_{\ti{n}} + 1}
   z_1^{2 k_1 + 1} z_2^{2 k_2 + 1} \cd z_n^{2 k_n + 1}} 
\eea
be the multi-loop generating function of these connected Green functions.  
This may be paraphrased as \cite{ajm}
\beq
   \lefteqn{\om_s (\z; z_1, z_2, \ld, z_n) = 
   \prod_{k=1}^n \le\lb - \sum_{j=1}^{\ift} \frac{2j}{z_k^{2j+1}} 
   \le( \frac{\partial}{\partial g_j} \ri)_{\b} \ri. } \nn \\
   & & \le. + \frac{\b}{(z_k^2 - \b)^{\frac{3}{2}}}
   \le( \frac{1}{W' (\b)} \frac{\partial}{\partial \b} \ri)_{g_j} \ri\rb
   \le( \frac{P_s (\z)}{\sqrt{\z^2 - \b}} \ri), 
\la{2.9a}
\eeq
where $\le( \frac{\partial}{\partial g_j} \ri)_{\b}$ is the partial 
differentiation operator with respect to $g_j$ with $\b$ held fixed, 
$\le( \frac{\partial}{\partial \b} \ri)_{g_j}$ is the partial
differentiation operator with respect to $\b$ with $g_1$, $g_2$, $g_3$, 
\ld, and so on held fixed, and $W (\b)$ was defined in Eq.~(\ref{2.5a}).  
According to Ref.~\cite{ajm},
\bea
   & & \le\lb - \sum_{j=1}^{\ift} \frac{2j}{z_k^{2j+1}} 
   \le( \frac{\partial}{\partial g_j} \ri)_{\b} 
   + \frac{\b}{(z_k^2 - \b)^{\frac{3}{2}}}
   \le( \frac{1}{W' (\b)} \frac{\partial}{\partial \b} \ri)_{g_j} \ri\rb 
   \le\lb \frac{1}{W' (\b)} \frac{\partial}{\partial \b} \ri\rb^n
   \frac{h(\b)}{W' (\b)} \\
   & & = \le\lb \frac{1}{W' (\b)} \frac{\partial}{\partial \b} 
   \ri\rb^{n+1} \frac{h(\b)}{W' (\b)} 
   \frac{\b}{(z_k^2 - \b)^{\frac{3}{2}}} 
\eea
if $n$ is a non-negative integer and $h (\b)$ is a function which depends 
only on $\b$ but not $g_1$, $g_2$, $g_3$, \ld, and so on.  As a result,
\bea
   \lefteqn{\om_s (\z; z_1, z_2, \ld, z_n) =} \\ 
   & & \le\lb \frac{1}{W' (\b)} \frac{\partial}{\partial \b} \ri\rb^{n-1}
   \le\lb \frac{\d_{sb} - \sum_{k=1}^{\ift} 
   \frac{(2k-1)!}{4^k k! (k-1)!} \b^k}{W' (\b) (\z^2 - \b)^{\frac{3}{2}}}
   \prod_{k=1}^n \frac{\b}{(z_k^2 - \b)^{\frac{3}{2}}} \ri\rb 
\eea
for any positive value of $n$.

In addition, Eq.~(\ref{2.9}) implies
\beq
   \om_s (\z_1, \z_2) = - \frac{2 \z_1 \z_2}{(\z_1^2 - \z_2^2)^2}
   + \frac{2 \z_1^2 \z_2^2 - \b (\z_1^2 + \z_2^2)}
   {(\z_1^2 - \z_2^2)^2 \sqrt{\z_1^2 - \b} \sqrt{\z_2^2 - \b}}. 
\la{2.10}
\eeq
Note that $\om_s (\z_1, \z_2)$ is independent of whether the model is
bosonic or fermionic and is independent of $c_1$, $c_2$, $c_3$, \ld, and
so on.  Thus we conclude from Eqs.~(\ref{2.10}) and (\ref{2.9}) that
\[ \om_s (\z_1, \z_2, \ld, \z_{\ti{n}}; z_1, z_2, \ld, z_n) = 0 \]
if $\ti{n} \geq 3$.  In other words,
\[ G_s (p_1, p_2, \ld, p_{\ti{n}}; k_1, k_2, \ld, k_n) = 0 \]
if $\ti{n} \geq 3$.  In terms of string worldsheet, this means that there
can be only two boundaries which are invariant under parity 
transformation.  Note also that Eq.~(\ref{2.10}) differs from the 
two-loop correlator of any complex matrix model by a factor of 4 only
\cite{ajm}.  Since $\om_s (\z_1, \z_2)$ depends on $g_1$, $g_2$, $g_3$,
\ld, and so on indirectly via $\b$ only, we could apply a formula similar
to Eq.~(\ref{2.9a}) to obtain other generating functions:
\beq 
   \lefteqn{\om_s (\z_1, \z_2; z_1, z_2, \ld, z_n) = } \nn \\
   & & \le\lb \frac{1}{W' (\b)} \frac{\partial}{\partial \b} \ri\rb^{n-1}
   \le\lb \frac{1}{2 \b W' (\b)} \frac{\b}{(\z_1^2 - \b)^{\frac{3}{2}}}
   \frac{\b}{(\z_2^2 - \b)^{\frac{3}{2}}}   
   \prod_{k=1}^n \frac{\b}{(z_k^2 - \b)^{\frac{3}{2}}} \ri\rb 
\la{2.11}
\eeq
for any positive value of $n$.  These multiloop generating functions differ 
from those of complex matrix models merely by constant factors of $2^{n+2}$ 
\cite{ajm}.  They are basically symmetry factors of the Feynman diagrams.

\subsection{Multicritical point}
\la{s2.3}

Following Ref.~\cite{kazakov}, we approach the $m$-th multicritical point by 
fine-tuning the coupling constants in such a way that there exists a 
real number $z_c$ which satisfies
\[ W (\b_*) = W^{(k)} (\b_*) = 0 \]
for $k = 1$, 2, \ld, and $m-1$, and
\[ W^{(m)} (\b_*) \neq 0. \]
Then
\[ W (\b) \simeq - \g (\b - \b_*)^m, \]
where $\g$ is a complex constant, for $\b$ close to $z_*$.  Let
\[ \z_i^2 = \b_* + a \ps_i, \; z_i^2 = \b_* + a \pi_i, \; \mbox{and} \; 
   \b = \b_* - a \sqrt{\La} \]
for any positive integer $i$, where $a$ is the cut-off length, $\La$ is the 
renormalised bulk cosmological constant, and $\pi_i$ and $\ps_i$ are 
renormalised boundary cosmological constants for any value of $i$.  Then
\[ \frac{1}{W' (\b)} \frac{\partial}{\partial \b} =
   - \frac{2}{m \g (-a)^m \La^{\frac{m-2}{2}}} 
   \frac{\partial}{\partial \La} \]
and we may conclude that the renormalised tree-level one-loop amplitude is
\[ \sqrt{a} \om_s (\ps_1) = \frac{2 \d_{sb} +
   \sum_{n=1}^{\ift} \b_*^n c_n \le\lb 2 \sum_{k=1}^n
   \frac{(2k-2)!}{4^k k! (k-1)!} - 1 \ri\rb}
   {(\ps_1 + \sqrt{\La})^{\frac{1}{2}}}, \]
and the renormalised tree-level multi-loop amplitudes are
\bea 
   \lefteqn{a^{(m + \frac{3}{2}) n + \frac{1}{2}} 
   \om_s (\ps_1; \pi_1, \pi_2, \ld, \pi_n; \La) =} \\
   & & \frac{(-1)^{mn + n + 1} 2^{n-2} \b_*^n \le\lb 1 - \sum_{k=1}^{\ift} 
   \frac{(2k-1)!}{4^k k! (k-1)!} \b_*^k \ri\rb} {m^n \g^n} \\
   & & \cdot \le( \frac{1}{\La^{\frac{m}{2} - \d_{sb}}} 
   \frac{\partial}{\partial \La} \ri)^{n-1}
   \le\lb \frac{1}{\La^{\frac{m-1}{2}}} 
   \frac{1}{(\ps_1 + \sqrt{\La})^{\frac{3}{2}}}   
   \prod_{k=1}^n \frac{1}{(\pi_i + \sqrt{\La})^{\frac{3}{2}}} \ri\rb
\eea
for $n \geq 1$, 
\[ a^2 \om_s (\ps_1, \ps_2) = - \frac{2 \b_*}{(\ps_1 - \ps_2)^2}
   + \frac{(\ps_1 + \ps_2 + 2 \sqrt{\La}) \b_*}
   {(\ps_1 - \ps_2)^2 (\ps_1 + \sqrt{\La})^{\frac{1}{2}}
   (\ps_2 + \sqrt{\La})^{\frac{1}{2}}} \]
and
\bea
   \lefteqn{a^{(m + \frac{3}{2}) n + 2}
   \om_s (\ps_1, \ps_2; \pi_1, \pi_2, \ld, \pi_n; \La) =} \\
   & & \frac{(-1)^{mn + n} \b_*^{n+1}}
   {4 m^n \g^n} 
   \le( \frac{1}{\La^{\frac{m}{2} - 1}} \frac{\partial}{\partial \La}
   \ri)^n \\
   & & \cdot \le\lb \frac{1}{\La^{\frac{m-1}{2}} 
   (\ps_1 + \sqrt{\La})^{\frac{3}{2}} (\ps_2 + \sqrt{\La})^{\frac{3}{2}}}
   \prod_{k=1}^n \frac{1}{(\pi_i + \sqrt{\La})^{\frac{3}{2}}} \ri\rb
\eea
for $n \geq 1$.

\section{Multicritical models of ${\mathbb R}^2 / {\mathbb Z}_2$} 
\la{s3}

Let us turn our attention to multicritical models of the quantum orbifold
${\mathbb R}^2 / {\mathbb Z}_2$.  The action of the bosonic version is
\bea
   \lefteqn{\ti{S}_b := - N_m \sqrt{N_v} \sum_{\m = 1}^{N_v} 
   \Tr M^{\da}_{\m} M_{\m}} \\
   & & - \frac{N_m^2 (g_1 - 1)}{2} \sum_{\m_1, \m_2 = 1}^{N_v} 
   \Tr \le( M^{\da}_{\m_1} M_{\m_2} \ri) 
   \Tr \le( M_{\m_2} M^{\da}_{\m_1} \ri) \\
   & & - N_m \sum_{n=1}^{\ift} c_n
   \sum_{\m_1, \m_2, \ld, \m_{2n} = 1}^{N_v} \Tr \le(
   M^{\da}_{\m_1} M_{\m_2} M^{\da}_{\m_3} M_{\m_4} \cd
   M^{\da}_{\m_{2n-1}} M_{\m_{2n}} \ri. \\
   & & \le. \cdot M_{\m_{2n}} M^{\da}_{\m_{2n-1}}
   M_{\m_{2n-2}} M^{\da}_{\m_{2n-3}} \cd M_{\m_2} M^{\da}_{\m_1} \ri) \\
   & & - N_m^2 \sum_{n=2}^{\ift} \frac{g_n}{2n}
   \sum_{\m_1, \m_2, \ld, \m_{2n} = 1}^{N_v} \Tr \le(
   M^{\da}_{\m_1} M_{\m_2} M^{\da}_{\m_3} M_{\m_4} \cd
   M^{\da}_{\m_{2n-1}} M_{\m_{2n}} \ri) \\
   & & \cdot \Tr \le( M_{\m_{2n}} M^{\da}_{\m_{2n-1}}
   M_{\m_{2n-2}} M^{\da}_{\m_{2n-3}} \cd M_{\m_2} M^{\da}_{\m_1} \ri),
\eea
whereas the action of the fermionic version is \cite{0303086}
\bea
   \lefteqn{\ti{S}_f := - N_m \sqrt{N_v} \sum_{\m = 1}^{N_v} 
   \Tr \bar{\Ps}_{\m} \Ps_{\m}} \\
   & & - \frac{N_m^2 (g_1 - 1)}{2} \sum_{\m_1, \m_2 = 1}^{N_v} 
   \Tr \le( \bar{\Ps}_{\m_1} \Ps_{\m_2} \ri) 
   \Tr \le( \Ps_{\m_2} \bar{\Ps}_{\m_1} \ri) \\
   & & - N_m \sum_{n=1}^{\ift} c_n
   \sum_{\m_1, \m_2, \ld, \m_{2n} = 1}^{N_v} \Tr \le(
   \bar{\Ps}_{\m_1} \Ps_{\m_2} \bar{\Ps}_{\m_3} \Ps_{\m_4} \cd
   \bar{\Ps}_{\m_{2n-1}} \Ps_{\m_{2n}} \ri. \\
   & & \le. \cdot \Ps_{\m_{2n}} \bar{\Ps}_{\m_{2n-1}}
   \Ps_{\m_{2n-2}} \bar{\Ps}_{\m_{2n-3}} \cd \Ps_{\m_2} \bar{\Ps}_{\m_1} \ri) 
   \\
   & & - N_m^2 \sum_{n=2}^{\ift} \frac{g_n}{2n}
   \sum_{\m_1, \m_2, \ld, \m_{2n} = 1}^{N_v} \Tr \le(
   \bar{\Ps}_{\m_1} \Ps_{\m_2} \bar{\Ps}_{\m_3} \Ps_{\m_4} \cd
   \bar{\Ps}_{\m_{2n-1}} \Ps_{\m_{2n}} \ri) \\
   & & \cdot \Tr \le( \Ps_{\m_{2n}} \bar{\Ps}_{\m_{2n-1}}
   \Ps_{\m_{2n-2}} \bar{\Ps}_{\m_{2n-3}} \cd \Ps_{\m_2} \bar{\Ps}_{\m_1} \ri).
\eea
Note that the second terms in these actions may be represented by a pair of
Feynman propagators.  The partition functions of the bosonic and fermionic 
models are defined as in Eqs.~(\ref{2.3}) and (\ref{2.1a}), respectively, with 
$Z$ replaced with $\ti{Z}$ and $S$ with $\ti{S}$.  For the bosonic model, the 
connected Green functions which we would like to study take the form
\beq
   \lefteqn{\ti{G}_b (p_1, p_2, \ld, p_{\ti{n}}; k_1, k_2, \ld, k_n) := 
   \lim_{N_m \rar \ift} \lim_{N_v \rar \ift} N_m^{\ti{n} + 2n - 2}} \nn \\ 
   & & \sum_{\n_{1, 1}, \n_{2, 1}, \ld, \n_{2 p_1, 1} = 1}^{N_v}
   \sum_{\n_{1, 2}, \n_{2, 2}, \ld, \n_{2 p_2, 2} = 1}^{N_v} \cd
   \sum_{\n_{1, \ti{n}}, \n_{2, \ti{n}}, \ld, \n_{2 p_{\ti{n}}, 
   \ti{n}} = 1}^{N_v} \nn \\ 
   & & \sum_{\m_{1, 1}, \m_{2, 1}, \ld, \m_{2 k_1, 1} = 1}^{N_v}
   \sum_{\m_{1, 2}, \m_{2, 2}, \ld, \m_{2 k_2, 2} = 1}^{N_v} \cd
   \sum_{\m_{1, n}, \m_{2, n}, \ld, \m_{2 k_n, n} = 1}^{N_v} \nn \\
   & & \le\langle \prod_{j=1}^{\ti{n}}
   \Tr \le( M^{\da}_{\n_{1, j}} M_{\n_{2, j}} M^{\da}_{\n_{3, j}} 
   M_{\n_{4, j}} \cd M^{\da}_{\n_{2p_j - 1, j}} M_{\n_{2p_j, j}} \ri. \ri. 
   \nn \\
   & & \le. \cdot M_{\n_{2p_j, j}} M^{\da}_{\n_{2p_j - 1, j}} 
   M_{\n_{2p_j - 2, j}} M^{\da}_{\n_{2p_j - 3, j}} \cd M_{\n_{2, j}} 
   M^{\da}_{\n_{1, j}} \ri) \nn \\
   & & \cdot \prod_{i=1}^n
   \Tr \le( M^{\da}_{\m_{1, i}} M_{\m_{2, i}} M^{\da}_{\m_{3, i}}
   M_{\m_{4, i}} \cd M^{\da}_{\m_{2k_i - 1, i}} M_{\m_{2k_i, i}} \ri) \nn \\ 
   & & \cdot \le. \Tr \le( M_{\m_{2k_i, i}} M^{\da}_{\m_{2k_i - 1, i}}
   M_{\m_{2k_i - 2, i}} M^{\da}_{\m_{2k_i - 3, i}} \cd M_{\m_{2, i}} 
   M^{\da}_{\m_{1, i}} \ri) \ri\rangle_{{\rm conn}, \ti{S}_b},
\la{3.1} 
\eeq
where $n$ is any non-negative integer, $\ti{n}$ is any positive integer, and
$p_1$, $p_2$, \ld, $p_{\ti{n}}$, $k_1$, $k_2$, \ld, and $k_n$ are any 
positive integers; for the fermionic model, the connected Green functions 
which we would like to study also take the form in Eq.~(\ref{3.1}) with $M$, 
$M^{\da}$, and $\ti{S}_b$ replaced with $\Ps$, $\bar{\Ps}$, and $\ti{S}_f$, 
respectively.  A Feynman diagram representing a term in a connected Green 
function is depicted in Fig.~\ref{f2}.

\begin{figure}
\centering
\includegraphics[width=2.9in]{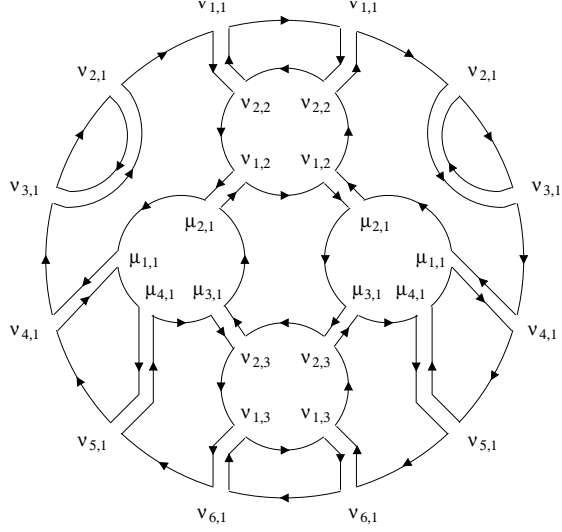}
\caption{\em A Feynman diagram of $\ti{S}_b (3, 1, 1; 2)$.}
\la{f2}
\end{figure}

\subsection{Schwinger--Dyson equations}
\la{s3.1}

To evaluate the connected Green functions at the double large-$N$ limit, 
let us start with the trivial equation
\beq
   \lefteqn{\lim_{N_m \rar \ift} \lim_{N_v \rar \ift}
   \frac{1}{N_m^2 \sqrt{N_v} Z'_b (N_m, N_v)}} \nn \\
   & & \cdot \sum_{i,j=1}^{N_m} 
   \sum_{\a_0 = 1}^{N_v} \int dM^{\da}_1 dM_1 dM^{\da}_2 dM_2 \cd
   dM^{\da}_{N_v} dM_{N_v} \nn \\
   & & \frac{\partial}{\partial M^{\da}_{\a_0 ij}} \{
   \sum_{\a_1, \a_2, \ld, \a_{2n-1} = 1}^{N_v}
   \le( M^{\da}_{\a_{2n-1}} M_{\a_{2n-2}} \cd M^{\da}_{\a_1} \ri. \nn \\
   & & \cdot \le. M^{\da}_{\a_1} M_{\a_2} \cd M^{\da}_{\a_{2n-1}} 
   M_{\a_0} \ri)_{ij} \exp \ti{S}_b \} = 0
\la{3.11}
\eeq
for the bosonic model or
\beq
   \lefteqn{\lim_{N_m \rar \ift} \lim_{N_v \rar \ift}
   \frac{1}{N_m^2 \sqrt{N_v} Z'_f (N_m, N_v)} \sum_{i,j=1}^{N_m} 
   \sum_{\a_0 = 1}^{N_v} \int d\Ps_1 d\bar{\Ps}_1 d\Ps_2 d\bar{\Ps}_2 \cd
   d\Ps_{N_v} d\bar{\Ps}_{N_v} } \nn \\
   & & \frac{\partial}{\partial \bar{\Ps}_{\a_0 ij}} \{
   \sum_{\a_1, \a_2, \ld, \a_{2n-1} = 1}^{N_v}
   \le( \bar{\Ps}_{\a_{2n-1}} \Ps_{\a_{2n-2}} \cd \bar{\Ps}_{\a_1}  
   \bar{\Ps}_{\a_1} \Ps_{\a_2} \cd \bar{\Ps}_{\a_{2n-1}} \Ps_{\a_0} 
   \ri)_{ij} \nn \\
   & & \exp \ti{S}_f \} = 0
\la{3.12}
\eeq
for the fermionic model.  Both Eqs.~(\ref{3.11}) and (\ref{3.12}) lead to
the Schwinger--Dyson equation
\beq
   \lefteqn{\sum_{k=1}^n \ti{\ph} (n-k) \ti{G} (k-1)} \nn \\
   & & - \sum_{k=1}^{\ift} c_k \sum_{l=1}^k \ti{G} (l-1) \ti{G} (n+k-l) -
   \sum_{k=1}^{\ift} g_k \ti{G} (n+k-1) = 0, 
\la{3.13}
\eeq
where $n$ is any positiver integer, $\ti{G} (n)$ stands for $\ti{G}_b (n)$ 
or $\ti{G}_f (n)$, and $\ti{\ph} (n)$ was defined in Eq.~(\ref{2.3a}).  
Hence the connected Green functions of the bosonic model at the double 
large-$N$ limit are identical to those of the fermionic model.

Let 
\[ \ti{\om} (\z) := \sum_{n=1}^{\ift} \frac{\ti{G} (n)}{\z^{2n+1}} \]
be the spectral function of these matrix-vector models.  It then follows 
from Eq.~(\ref{3.13}) and the well-known expression for the spectral
function $\c (\z)$ of the ordinary Hermitian matrix model that
\beq
   \ti{\om} (\z) & = & \frac{2 Q_1 (\z)}
   {Q_2 (\z) + M(\z, \b) \sqrt{\z^2 - \b}}, 
\la{3.14} \\
   & = & \frac{2 Q_1 (\z) \le\lb Q_2 (\z) - M(\z, \b) \sqrt{\z^2 - \b} 
   \ri\rb}{Q_2^2 (\z) - M^2 (\z, \b) (\z^2 - \b)}
\la{3.14c}
\eeq
where $\b$ and $M (\z, \b)$ were defined in Eqs.~(\ref{2.5a}) and 
(\ref{2.5b}),
\beq
   Q_1 (\z) := 
   \sum_{k=1}^{\ift} \z^{2k-2} \sum_{l=0}^{\ift} g_{k+l} \ti{G} (l) 
   + \sum_{k=0}^{\ift} \z^{2k} \sum_{l=0}^{\ift} \ti{G} (l) 
   \sum_{m=0}^{\ift} \ti{G} (m) c_{k+l+m+1}, 
\la{3.14a} 
\eeq
and
\beq
   Q_2 (\z) := \sum_{k=1}^{\ift} g_k \z^{2k-1} +
   2 \sum_{k=0}^{\ift} \z^{2k+1} \sum_{l=0}^{\ift} \ti{G} (l) c_{k+l+1}. 
\la{3.14b}
\eeq
As usual, we assert that the values of the connected Green functions in 
Eqs.~(\ref{3.14a}) and (\ref{3.14b}) are determined by the requirement that 
$\ti{\om}$ be holomorphic on the whole complex plane except the branch cut 
$- \sqrt{\b} \leq \Re (\z) \leq \sqrt{\b}$ and $\Im (\z) = 0$.

\subsection{Some multicritical points}
\la{s3.2}

A convenient choice of the $m$-th multicritical point is to select a non-zero
value of $g_m$, adjust the values of $g_1$, $g_2$, \ld, and $g_{m-1}$ such 
that 
\[ M (\z) = g_m (\z^2 - \b)^{m-1}, \]
and adjust the values of $c_1$, $c_2$, \ld, and $c_m$ such that
\beq
   Q_2 (\z) = g_m \z (\z^2 - \b)^{m-1}. 
\la{3.15a}
\eeq
Moreover, $g_n = c_n = 0$ if $n > m \geq 2$.  It then follows from 
Eq.~(\ref{3.14}) that 
\beq
   \ti{\om} (\z) = \frac{2 Q_1 (\z) (\z - \sqrt{\z^2 - \b})}
   {g_m \b (\z^2 - \b)^{m-1}}.
\la{3.15}
\eeq
The holomorphic property of $\ti{\om} (\z)$ then dictates that the zeros of 
$Q_1 (\z)$ coincide with the zeros of the denominator on the right side of 
Eq.~(\ref{3.15}).  As a result, at the $m$-th multicritical point,
\beq
   Q_1 (\z) = A (\z^2 - \b)^{m-1}; 
\la{3.15b}
\eeq
the constant $A$ may be determined by the condition that
\[ \lim_{\z \rar \ift} \z \ti{\om} (\z) = 1. \]
This yields $A = g_m$.  As a result,
\[ \ti{\om} (\z) = \frac{2}{\b} \le( \z - \sqrt{\z^2 - \b} \ri) \]
at the $m$-th multicritical point.

A convenient way to approach the $m$-th multicritical point is to keep 
the ratios $g_i : g_j$, $g_i : c_j$, and $c_i: c_j$, where $i$ and $j$ are
positive integers less than or equal to $m$, fixed.  Then in Eq.~(\ref{3.14}),
only $\z$ and $\b$ deviates from their critical values $\sqrt{\b_*}$ and 
$\b_*$, respectively, whereas $g_k$, $c_k$, and $\ti{G} (k)$, where $k$ is any 
positive integer not larger than $m$, are fixed.  Let
\[ \z^2 = \b_* + a \pi \; \mbox{and} \; \b = \b_* - a \sqrt{\La}, \]
where $a$ is the cut-off length, and $\pi$ and $\La$ are the boundary and bulk 
cosmological constants, respectively.  Then $Q_1 (\z)$, $Q_2 (\z)$, and 
$M (\z)$ are of order $a^{m-1}$, whereas $\sqrt{\z^2 - \b}$ is of order 
$\sqrt{a}$.  Recall that $M (\z) \sqrt{\z^2 - \b}$ is, up to a proportionality 
constant, also the singular part of the spectral function of ordinary Hermitian 
matrix models.  Hence we conclude from Eqs.~(\ref{3.14c}), (\ref{3.15a}), and
(\ref{3.15b}) that we may multiply $\ti{\om}$ by $\sqrt{a}$ to obtain the 
renormalised tree-level one-loop amplitude which, up to a constant factor, is
\[ \frac{1}{\pi^{m-1}} \le( \ba{c} \mbox{renormalised 1-loop amplitude of} \\
   \mbox{an ordinary Hermitian matrix model} \\ 
   \mbox{at the $m$-th multicritical point} \ea \ri). \]

\section{Conclusion and Outlook}
\la{s4}

We may study quantum orbifold geometry by means of bosonic or fermionic
matrix-vector models.  As for the quantum orbifold ${\mathbb R}^2 / 
{\mathbb Z}_2 \times {\mathbb Z}_2$, the bosonic model differs from the 
fermionic model in the sense that Feynman diagrams with no $c_i$-vertices,
where $i$ is any positive integer, contribute to the Green functions of the 
bosonic model only; they have no contribution to those of the fermionic model.  
If in an orbifolded worldsheet there is only one boundary which is invariant 
under parity transformation, then its multiloop amplitude is significantly 
different from that of an ordinary worldsheet.  Nonetheless, if there are two 
boundaries which are invariant under parity transformation, then its 
multiloop amplitude is the same as that of an ordinary worldsheet up to a 
symmetry factor.

As for the quantum orbifold ${\mathbb R}^2 / {\mathbb Z}_2$, the bosonic and
fermionic models are equivalent to each other at the double large-$N$ limit.
The renormalised tree-level one-loop amplitude at an $m$-th multicritical
point differs from that of an ordinary Hermitian matrix model by a factor 
inversely proportional to $\pi^{m-1}$.  Nevertheless, it may be possible to 
identify other $m$-th multicritical points at which the quantum orbifold may behave 
differently.  It would also be of interest to obtain more explicity expressions 
for higher loop amplitudes of this quantum orbifold.  Furthermore, exploring the 
double-scaling limit of these matrix-vector models would give us valuable 
information on the non-perturbative behavior of unoriented string theory.

\pagebreak

\noindent \Large{\bf \hskip .2pc Acknowledgment}
\vskip 1pc

\normalsize

\noindent
I thank A. Nica and R. Szabo for discussions.  This work is partially 
supported by the Pure Mathematics Department at the University of Waterloo.


\begin{thebibliography}{99}
\bibitem{dgz} P. Di Francesco, P. Ginsparg, and J. Zinn-Justin, {\em 2-D 
   gravity and random matrices}, Phys. Rept. {\bf 254} (1995) 1---133 
   {\tt \lb hep-th/9306153\rb }.
\bibitem{mv} J. McGreevy and H. Verlinde, {\em Strings from tachyons: the
   $c$ = 1 matrix reloaded} {\tt \lb hep-th/0304224\rb }.
\bibitem{dkkmms} M. R. Douglas, I. R. Klebanov, D. Kutasov, J. Maldacena,
   E. Martinec, and N. Seiberg, {\em A new hat for the $c$ = 1 matrix
   model} {\tt \lb hep-th/0307195\rb }.
\bibitem{ps} G. Pradisi and A. Sagnotti, {\em Open string orbifolds}, 
   Phys. Lett. {\bf B 216} (1989) 59---67.
\bibitem{horava} P. Ho\v{r}ava, {\em Strings on world-sheet 
   orbifolds}, Nucl. Phys. {\bf B 327} (1989) 461---484.
\bibitem{schwarz} J. H. Schwarz, in {\em Strings, branes and gravity:
   lecture notes TASI 99 Boulder, Colorado, USA 31 May --- 25 June 1999},
   ed. J. A. Harvey et al. (World Scientific, 1999) pp.809---846 {\tt
   \lb hep-th/9908144\rb }.
\bibitem{bgn} P. Biane, F. Goodman, and A. Nica, {\em Non-crossing
   cumulants of type~B}, {\tt math.OA/0206167}.
\bibitem{0303086} C.-W. H. Lee, {\em Noncommutative probability, matrix 
   models, and quantum orbifold geometry}, JHEP {\bf 06} (2003) 044 {\tt \lb
   hep-th/0303086\rb }.
\bibitem{ajm} J. Ambj{\o}rn, J. Jurkiewicz, and Yu. M. Makeenko, {\em 
   Multiloop correlators for two-dimensional quantum gravity}, Phys. Lett.
   {\bf B 251} (1990) 517---524
\bibitem{kazakov} V. A. Kazakov, {\em The appearance of matter fields from 
   quantum fluctuations of 2D-gravity}, Mod. Phys. Lett. {\bf A} (1989) 
   2125---2139.
\end{thebibliography}
\end{document}